\documentclass{article}

\usepackage{url} 
\usepackage{hyperref}

\usepackage[utf8]{inputenc}  
\usepackage[T1]{fontenc}     
\usepackage{hyperref}        
\usepackage{url}             
\usepackage{graphicx}        
\usepackage{xcolor}          
\usepackage{lipsum}          

\usepackage{amsmath, amssymb}  
\usepackage{amsfonts}          
\usepackage{mathtools}         
\usepackage{bm}                

\usepackage{booktabs}          
\usepackage{multirow}          
\usepackage{array, longtable, tabularx}  
\usepackage{adjustbox}         
\usepackage{epsfig}            
\usepackage{lscape}            

\usepackage{algorithm}
\usepackage{algorithmic}

\usepackage{gensymb}           
\usepackage{scalerel, stackengine}  
\stackMath

\usepackage{natbib}            
\usepackage{doi}               

\usepackage{microtype}         
\usepackage{nicefrac}          
\usepackage{arxiv}             

\newcommand{\argmax}{\arg\!\max}

\newcommand*{\tran}{^{\mkern-1.5mu\mathsf{T}}}

\newcommand\reallywidehat[1]{%
    \savestack{\tmpbox}{\stretchto{%
        \scaleto{%
            \scalerel*[\widthof{\ensuremath{#1}}]{\kern.1pt\mathchar"0362\kern.1pt}%
            {\rule{0ex}{\textheight}}
        }{\textheight}%
    }{2.4ex}}%
    \stackon[-6.9pt]{#1}{\tmpbox}%
}
\newcommand{\comment}[1]{}

\author{%
  \begin{minipage}{.45\textwidth}
    \centering
    \href{https://orcid.org/0000-0001-5935-6018}{\includegraphics[scale=0.06]{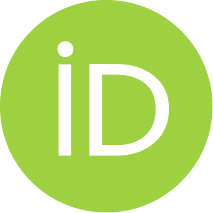}\hspace{1mm}Lanqiu Yao} \\
Magma Capital Management LLC \\
24 E Washington St, Suite 875 \\
Chicago, IL 60602\\
    \texttt{ly1192@nyu.edu}
  \end{minipage}%
  \hspace{.1\textwidth}
  \begin{minipage}{.45\textwidth}
    \centering
    \href{https://orcid.org/0000-0002-3964-5899}{\includegraphics[scale=0.06]{orcid.pdf}\hspace{1mm}Thaddeus Tarpey} \\
    Division of Biostatistics, \\ 
    Department of Population Health,\\
    NYU Grossman School of Medicine \\
    New York, NY 10016 \\
    \texttt{Thaddeus.Tarpey@nyulangone.org}
  \end{minipage}
}
\date{\vspace{-5ex}}
   
\title{Trajectory-Based Individualized Treatment Rules}

\author{ 
    \href{https://orcid.org/0000-0001-5935-6018}{\includegraphics[scale=0.06]{orcid.pdf}\hspace{1mm}Lanqiu Yao} \\
    Magma Capital Management LLC \\
    24 E Washington St, Suite 875 \\
    Chicago, IL 60602\\
    \texttt{ly1192@nyu.edu} 
    \and
    \href{https://orcid.org/0000-0002-3964-5899}{\includegraphics[scale=0.06]{orcid.pdf}\hspace{1mm}Thaddeus Tarpey} \\
    Division of Biostatistics, \\ 
    Department of Population Health,\\
    NYU Grossman School of Medicine \\
    New York, NY 10016 \\
    \texttt{Thaddeus.Tarpey@nyulangone.org}
}

\renewcommand{\shorttitle}{\textit{arXiv} Template}

\hypersetup{
pdftitle={A template for the arxiv style},
pdfsubject={q-bio.NC, q-bio.QM},
pdfauthor={Lanqiu Yao, Thaddeus Tarpey},
pdfkeywords={},
}

\begin{document}
\maketitle

\begin{abstract}
A core component of precision medicine research involves optimizing individualized treatment rules (ITRs) based on patient characteristics. 
Many studies used to estimate ITRs are longitudinal in nature, collecting outcomes over time. 
Yet, to date, methods developed to estimate ITRs often ignore the longitudinal structure of the data. 
Information available from the longitudinal nature of the data can be especially useful in mental health studies.
Although treatment means might appear similar, understanding the trajectory of outcomes over time can reveal important differences between treatments and placebo effects.
This longitudinal perspective is especially beneficial in mental health research, where subtle shifts in outcome patterns can hold significant implications.
Despite numerous studies involving the collection of outcome data across various time points, most precision medicine methods used to develop ITRs overlook the information available from the longitudinal structure. 
The prevalence of missing data in such studies exacerbates the issue, as neglecting the longitudinal nature of the data can significantly impair the effectiveness of treatment rules.
This paper develops a powerful longitudinal trajectory-based ITR construction method that incorporates baseline variables, via a single-index or biosignature, into the modeling of longitudinal outcomes. 
This trajectory-based ITR approach substantially minimizes the negative impact of missing data compared to more traditional ITR approaches.
The approach is illustrated through simulation studies and a clinical trial for depression, contrasting it with more traditional ITRs that ignore longitudinal information.
\end{abstract}

\section{Introduction}\label{intro} 

A central objective of precision medicine is to utilize patient baseline characteristics to establish optimal treatment decision rules. 
This approach aids in identifying baseline variables that predict patient treatment responses as well as responses due to placebo effects \citep{tarpey2016stratified}. 
The development of \textit{individualized treatment rules} (ITRs), tailored to each patient's unique characteristics, represents a significant advancement over the traditional ``one-size-fits-all'' approach, which often fails to account for the diverse responses observed in medical research \citep[e.g.,][]{kosorok2019precision}. 
In many settings, such as depression studies, the influence of individual predictors on treatment outcomes is minimal, prompting the need for developing composite modifiers.

Several ITR approaches have been proposed, such as Q-learning ~\citep{watkins1989learning}, where ``Q'' signifies ``quality''.
This strategy involves modeling Q-functions in various ways, including parametric, semiparametric, and nonparametric approaches ~\citep{zhao2009reinforcement}. 
Advantage learning (A-learning), introduced by ~\cite{murphy2003optimal}, presents an alternative to Q-learning and is known for its robustness against model misspecification ~\citep{schulte2014q}.
Regression-based methods have also been explored to develop ITRs \citep[e.g.,][]{petkova2020optimising, tian2014simple, henderson2010regret, park2020single, park2021constrained, song2017semiparametric, fan2017concordance}, including those capable of incorporating functional predictors \citep[e.g.,][]{Ciarleglio.2015, CiarleglioEtAl2016, CiarleglioEtAl2018, ParkEtAl2021}. 
Many of these methods are grounded in single-index models that utilize a \textit{Generated Effect Modifier} (GEM) \citep{petkova2017generated}, essentially a composite defined as a linear combination of baseline patient characteristics, as a predictor in regression models.
\citet{park2020single} introduced a robust \textit{single-index model with multiple links} (SIMML) functions as a generalization of the GEM that links the outcome to a smooth nonparametric GEM.  
\cite{song2017semiparametric} proposed a semiparametric additive single-index model that includes a link function to interact with treatment and baseline features.
Machine learning techniques have also been increasingly applied in developing ITRs, including Outcome Weighted Learning (OWL) \citep[e.g.,][]{zhao2012estimating} using Support Vector Machines (SVM), tree-based methods \citep[e.g.,][]{laber2015tree}, and neural networks and evolutionary algorithms~\citep{uddin2019artificial}.

Most studies used to develop ITRs collect data longitudinally, but most ITR methods do not utilize information from this longitudinal nature of the data.
Instead, they rely solely on a change score outcome (the difference between the last and first measurements) to develop the ITRs. 
The longitudinal aspect of the data holds significant potential for enhancing the accuracy of treatment decisions. 
For instance, patient responses can vary, with some showing immediate effects, potentially due to placebo responses, and others exhibiting delayed reactions.
Additionally, the varying mechanisms of action across different treatments can lead to distinct patterns in the progression of outcome trajectories. 
In such cases, although two treatments may appear similar when evaluated using a simple change score, baseline covariates could reveal diverse patterns of treatment responses over time.

Incorporating information from response trajectories, which are essentially outcome functions of time, presents a challenge due to the lack of a natural order among these curves. 
The absence of a natural ordering of curves makes it difficult to rank treatments from best to worst for individual patients. 
This paper derives a scalar measure from outcome trajectories that can be used to optimize Individualized Treatment Rules (ITRs) based on the longitudinal characteristics present in the data.

Extracting a useful scalar summary from longitudinal data has been explored previously \citep[e.g.,][]{wishart1938growth}. 
A common method involves calculating the slope from a linear fit to the data \citep{vossoughi2012summary}, but this approach may not be ideal for nonlinear curves. 
An alternative method involves determining a summary measure that reflects the average rate of improvement or the {\em average tangent slope} (ATS) of the curve \citep{tarpey2021extracting}. 
\cite{yao2022single} introduced a novel longitudinal single-index model, known as LS-KLD (Longitudinal Single-index model based on Kullback-Leibler Divergence), which effectively distinguishes between longitudinal trajectories across different treatment groups. 
This model utilizes the ATS to inform ITRs. 
Its primary criterion for optimization is the maximization of the KLD between the outcome distributions of various treatments. 
Consequently, LS-KLD demonstrates enhanced performance compared to conventional ITRs, such as a higher rate of correct decisions. 
This improvement is notable compared to commonly used methods like ANCOVA or methods based on the slope of a fitted straight line. 
This paper offers alternatives to the LS-KLD approach for extracting a scalar summary from curve trajectories that can achieve better performance by directly estimating the ATS. 
Another challenge frequently encountered in longitudinal studies is missing data, such as missed follow-up assessments or participant dropout. 
The trajectory-based ITRs considered in this paper can minimize the negative impact of missing data compared to traditional scalar-based ITRs.

In Section \ref{approach}, the trajectory-based single-index modeling approaches for constructing ITRs are introduced. 
These methods are compared against traditional ITR approaches that ignore information from the longitudinal data structures. 
This comparative analysis is conducted through simulation experiments detailed in Section \ref{simsection} and within the context of a depression randomized clinical trial described in Section \ref{embarc}. 
The paper concludes with a discussion in Section \ref{discussion}.

\section{Notation and Approach}\label{approach}

Consider data from a randomized longitudinal study comparing $K$ groups (e.g., $K$ medical treatments). 
We observe outcomes for subjects indexed by $i$, within groups indexed by $k$ ($k= 1,.., K$), and at multiple time points indexed by $j$ for $j\in\{1,..., m_{ik}\}$.

Let $\tilde{y}_{ijk}$ denote the observed outcome for the $i$th subject in the $k$th group at time $t_{ijk}$, where $t_{ijk}$ represents the measurement times common across observations, often fixed in designed studies as $t^*_j$. 
The vector of observed outcomes for subject $i$ in group $k$ is denoted by $\tilde{\bm{Y}}_{ik} = (\tilde{y}_{i1k}, \tilde{y}_{i2k}, ..., \tilde{y}_{im_{ik}k})\tran$.
We assume the observed outcomes $\tilde{y}_{ijk}$ are given by the equation:
\begin{equation}\label{tmp1}
\tilde{y}_{ijk} = y_{ik}(t) + \epsilon_{ijk}, \quad \text{with } t = t_{ijk}.
\end{equation}
Here, $y_{ik}(t)$ represents a continuous-time process modeling the functional outcome, and $\epsilon_{ijk}$ denotes a random error.

Baseline covariates for each subject in each group are denoted as $\bm{x}_{ik}\tran= (x_{ik1}, ..., x_{ikp})\in \mathbb{R}^p$. 
These covariates are used to define a linear generated effect modifier (or biosignature) $\bm{\alpha}\tran \bm{x}_{ik}$, where $\bm{\alpha}\in \mathbb{R}^p$. 
To ensure identifiability of $\bm{\alpha}$, we impose the constraint $||\bm{\alpha}||_2 = 1$ with the first element being positive.

The mean outcome, influenced by both time and the biosignature, is modeled as:
\begin{equation}\label{tmp2}
\mu_k(t; \bm{x}) = E(y_{ik}(t; \bm{x})) = h_k(t, \bm{\alpha} \tran \bm{x}),
\end{equation}
where $h_k(\cdot)$ is a smooth link function. 
We approximate this function using a tensor product of basis functions:
\[
\mu_k(t; \bm{x}) \approx \bm{\eta}_k \tran \left[\bm{g}(t) \otimes \bm{a}(\bm{\alpha} \tran \bm{x})\right].
\]
Here, $\bm{g}(t) = [g_1(t), g_2(t), ..., g_{d_1}(t)]^{\mkern-1.5mu\mathsf{T}}$ represents basis functions of time and $\bm{a}(u) = [a_1(u), ..., a_{d_2}(u)]^{\mkern-1.5mu\mathsf{T}}$ represents basis functions for the biosignature $u = \bm{\alpha} \tran \bm{x}$.

The linear biosignature $\bm \alpha\tran \bm x_{ik}$ will be incorporated into an ITR by modeling the longitudinal outcome using the mixed-effects model:
\begin{equation}\label{lme1}
    \tilde{\bm Y}_{ik} \approx \bm X_{ik\bm \alpha} \bm \eta_k +  \bm Z_{ik} \bm b_{ik} + \bm \epsilon_{i,k},
\end{equation}
where $\bm{X}_{ik\bm{\alpha}} = \bm{G}_{ik} \otimes \big[\bm{a}(\bm{\alpha}\tran\bm{x}_{ik}) \big]^{\mkern-1.5mu\mathsf{T}}$, $\bm{G}_{ik}$ is an $m_{ik}\times d_1$ matrix with rows $\bm{g}(t_j)\tran$, and $\bm{b}_{ik}$ is a mean-zero vector of random effects with design matrix $\bm{Z}_{ik}$ dependent on time only.

\subsection{Extracting a Scalar Response: Average Tangent Slope (ATS)}\label{ATSsection}

In order to construct an ITR to assign one of $K$ treatments to an individual based on functional outcomes, it is necessary to derive an ordering (e.g., best to worst) among the potential functional outcomes for an individual under different treatments. 
Since there is no natural ordering in the space of real-valued functions, the goal is to find a suitable scalar summary of the functional outcomes so that an ordering of the outcomes can be obtained to define an ITR. 
For an individual with baseline covariates $\bm{x}$, a simple approach is to assign a treatment by comparing the change (or improvement) in the expected potential outcomes from baseline $t_1^\ast$ to the end of the study $t_m^\ast$:
\begin{equation}\label{meanchange}
\mu_k(t_m^\ast; \bm x)-\mu_k(t_1^\ast;\bm{x}),
\end{equation} 
across groups $k=1,\dots,K.$
If the potential outcomes at times $t_{1}^\ast$ and $t_m^\ast$ were available for an individual for all treatments, then 
the {\em change score} (CS), 
\begin{equation}\label{CS1}
CS_{ik} = \tilde{y}_{ik}(t_m^\ast) - \tilde{y}_{ik}(t_1^\ast),
\end{equation}
could be compared across groups to make a treatment decision.
However, although (\ref{CS1}) is conditionally unbiased (given $\bm{x}$) for (\ref{meanchange}), it uses only the first and last observations and ignores all other data and hence is noisy and inefficient.

Since the potential outcomes are generally not available, treatment decisions can be made by using estimates of model parameters in (\ref{lme1}) to predict the CS in (\ref{CS1}).  
The derivative $d \mu_k (t | \bm \alpha \tran \bm x)/dt$ is the tangent slope (or instantaneous rate of change) of the outcome curve at time $t$.
By the Fundamental Theorem of Calculus, 
the integral of $d \mu_k (t | \bm \alpha \tran \bm x)/dt$ from $t_1^\ast$ to $t_m^\ast$ equals the change score and if this is divided by the length of the time interval, 
we obtain the {\em average tangent slope} (ATS) or average rate of improvement,  which is the summary scalar measure we propose to use to compare curves for different groups.
Using our notation, the ATS can be expressed as
\begin{equation}
  \text{ATS}_k(u) :=  \frac{\mu_k(t_m^\ast, u) - \mu_k (t_1^\ast, u)}{t_m^\ast - t_1^\ast} 
   =  \frac{\bm \eta_k \tran \big[ (\bm{g}(t_m^\ast) - \bm{g}(t_1^\ast)) \otimes \bm a(u) \big]}{t_m^\ast -t_1^\ast},
\end{equation}
with $u = \bm{\alpha}\tran \bm{x}$. 
Using estimates of the ATS allows efficiency gains by employing all available longitudinal data.

\subsection{Linear Biosignature Construction}\label{construct}

In order to optimize an IRT using a linear biosignature $\bm{\alpha}\tran \bm{x}$ and the longitudinal information, the objective is to find the $\bm{\alpha}$ that best differentiates ATS for different groups. 
The criterion used here is to find $\bm{\alpha}$ that maximizes the expected squared differences between the ATS of different groups across the distribution of the baseline covariates $\bm{x}$:
\begin{equation}\label{alphadefn}
\reallywidehat{\bm{\alpha}} = \arg\max_{\bm{\alpha}} \int \big(\text{ATS}_{1} (s) - \text{ATS}_{2} (s) \big)^2 f_{\bm \alpha} (s) ds,
\end{equation}
where $f_{\bm \alpha} (s) $ is the density of $S= \bm{\alpha}\tran\bm{x}$
 (this can be generalized easily to $K>2$ groups). 
From our notation, the objective to be optimized (\ref{alphadefn}) can be expressed
\begin{eqnarray}\nonumber
      \int \Big(\text{ATS}_1(u) - \text{ATS}_2(u)\Big)^2 f_{\bm \alpha}(u) du 
      &=&
       \frac{1}{(t_m^* -t_1^*)^2}\int (\bm \eta_1 - \bm \eta_2) \tran \big[ (\bm G(t_m^*) - \bm G(t_1^*)) \otimes \bm a(u) \big] \times \\\label{chap2-eq1}
      &&\;\;\;\;\;\;\;\;\;  \big[ (\bm G(t_m^*) - \bm G(t_1^*))\tran \otimes \bm a(u) \tran \big]  (\bm \eta_1 - \bm \eta_2)f_{\bm \alpha}(u) du \\\nonumber
      &= & \frac{1}{(t_m^* -t_1^*)^2}\int \color{black}(\bm \eta_1 - \bm \eta_2) \tran \Big[ \bm G^* \otimes \bm a(u) \bm a(u)\tran \Big] (\bm \eta_1 - \bm \eta_2)f_{\bm \alpha}(u) \color{black} du,\\\nonumber
\end{eqnarray}
where $\bm G^* = (\bm G(t_m^*) - \bm G(t_1^*)) \otimes (\bm G(t_m^*) - \bm G(t_1^*))\tran$. 

If a flexible set of basis functions are used to fit model (\ref{lme1}), then  ``nonparametric'' estimators of the ATS can be obtained and we define the 
{\em nonparametric average tangent slope} (NPATS) estimator of $\bm{\alpha}$ 
by 
\begin{equation}\label{npats}
 \reallywidehat{\bm \alpha} ^\text{NPATS}=  \argmax_{\bm \alpha} \int \big(\widehat{\text{ATS}}_{1} (s) - \widehat{\text{ATS}}_{2} (s) \big)^2 \hat{f}_{\bm \alpha} (s) du.
\end{equation}
Many longitudinal randomized studies are generally short-term, typically lasting between 6 to 12 weeks. 
In such scenarios, straightforward functional forms, like quadratic curves, are often effective in closely approximating the trajectories of outcomes. 
Therefore, in many settings, using a parametric mixed-effects model,  generates well-performing ITRs.
Specifically, we consider a parametric special case of model (\ref{lme1}),
\begin{equation}\label{lme2}
    \tilde{\bm Y}_{ik} = \bm{G}_{ik} \big(\bm \beta_k +  \bm \alpha \tran \bm x_{ik} \bm \Gamma_k  \big) +   \bm{Z}_{ik} \bm b_{ik} + \bm \epsilon_{ik},
\end{equation}
where $\bm{\beta}_k$ is a group-specific coefficient vector of fixed time trend effects and
$\bm \Gamma_k$ is a fixed-effect group-interaction coefficient vector that models the linear biosignature's impact on the outcome;
$\bm b_{ik}$ is the subject-specific vector of random effects and 
$\bm \epsilon_{ik}$ is a vector of random errors.  
For ITR estimation,  the mixed-effect models (\ref{lme1}) and (\ref{lme2}) will specify $\bm b_{ik} \sim N(\bm 0, \bm D_k) \perp  \bm \epsilon_{ik} \sim N(0, \sigma_k^2 \bm I)$.

In the application for this paper, the outcome trajectories from most individuals are well-approximated by quadratic curves in which case $d_1=3$ and we set the rows of
$\bm{Z}_{ik}$ to be of the form $(1,t,t^2)$.
For the simpler model (\ref{lme2}), the ATS in group $k$ conditional on the linear biosignature $\bm{\alpha}\tran\bm{x}$ will be called the {\em parametric average tangent slope} (PATS)
and is given by
\begin{equation} \label{ats_k}
     \text{PATS}_k (\bm \alpha \tran \bm x) = 
      \frac{(\bm{g}(t_m^*) - \bm{g}(t_1^*))}{t_m^* - t_1^*} \Big( \bm \beta_k + \bm \alpha \tran \bm x \bm{\Gamma}_k \Big),
\end{equation}
and  the estimated linear biosignature  for defining an ITR for this simpler model 
will be denoted as $\reallywidehat{\bm \alpha}^\text{PATS}.$

\subsection{Estimation of the Linear Biosignature}\label{est}

If $\bm \alpha$ in (\ref{lme1}) is known and assuming the random effects and the random error are both normally distributed and independent of each other, then from standard results from linear mixed-effect models, a closed form expression for the fixed-effect coefficient $\bm{\eta}_k$ estimator in each group is
\begin{equation} \label{chap2-eq0}
\widehat{\bm \eta}_k = \big(\bm X_{\bm \alpha}^{\mkern-1.5mu\mathsf{T}}  \reallywidehat{\bm V}_k^{-1} \bm X_{\bm \alpha} \big)^{-1} \bm X_{ \bm \alpha} ^{\mkern-1.5mu\mathsf{T}} \reallywidehat{\bm V}_k^{-1} \tilde{\bm Y}_{k}, 
\end{equation}
where $\reallywidehat{\bm V}_k$ is the estimated covariance matrix for the outcome in group $k$, 
$\bm X_{\bm \alpha} = \Big(\bm X_{1k\bm \alpha}\tran, ..., \bm X_{{n_k}k\bm \alpha} \tran \Big)\tran$,
and $ \tilde{\bm Y}_k = \Big(\tilde{\bm Y}\tran _{1k},..., \tilde{\bm Y}\tran _{n_k k}\Big)\tran.$
Optimizing (\ref{npats}) can be done quickly using the Nelder-Mead Algorithm \citep{nelder1965simplex}. 
Here, we illustrate for $K=2$.
Given an $\bm \alpha$ value, we can evaluate $\reallywidehat{\bm \eta}_k$  using (\ref{chap2-eq0});
a candidate for $\reallywidehat{\bm \alpha}^\text{NPATS}$ can then be obtained from (\ref{npats}) using (\ref{chap2-eq1}) and an empirical distribution
for $\bm{\alpha}\tran\bm{x}$ to 
approximate the density  $\hat{f}_{\bm \alpha} (s)$.
Given that basis functions $\bm{g(}t)$ and $\bm a(u)$ have been selected, the estimation algorithm is summarized as: 
\begin{algorithm}[H]
\floatname{algorithm}{Algorithm}
\caption{Estimation of $\bm \alpha$ by maximizing the square difference of ATS}
\label{protocol1}
\begin{algorithmic}[1]
\STATE Initialize $\reallywidehat{\bm \alpha}^{\text{NPATS}(l)}, l=0$
\STATE Given $\reallywidehat{\bm \alpha}^{\text{NPATS}(l)}$, calculate the coefficients of the basis functions $\reallywidehat{\bm \eta}_k^{(l)}$ from (\ref{chap2-eq0})
\STATE Given $\reallywidehat{\bm \eta}_k^{(l)}$, update $\reallywidehat{\bm \alpha}^{\text{NPATS}(l+1)}$ based on (\ref{npats}) using (\ref{chap2-eq1}) 
\STATE Repeat steps 2 \& 3 until convergence.
\end{algorithmic}
\end{algorithm} 
The convergence criterion used for this algorithm was to iterate until 
the cosine similarity between $\reallywidehat{\bm \alpha}^{\text{NPATS}(l)}$ and $\reallywidehat{\bm \alpha}^{\text{NPATS}(l+1)}$ is greater than 0.99.
In our simulation illustrations in Section \ref{simsection},
the algorithm usually converges within 50 iterations.

\subsubsection{Parametric Average Tangent Slope Estimation of Linear Biosignature}\label{opt_ats}

For the
the  parametric modeling approach (\ref{lme2}), 
the criterion to estimate $\bm \alpha$ is formulated as:
\begin{equation}\label{catss}
\begin{aligned}
    \reallywidehat{\bm \alpha} ^\text{PATS}= & \argmax_{\bm \alpha} \int \big(\widehat{\text{ATS}}_{1} (s) - \widehat{\text{ATS}}_{2} (s) \big)^2 \hat{f}_{\bm \alpha} (s) du. \\
    = & \; c_1 
+ c_2  \reallywidehat{\bm \mu}_x \tran \bm \alpha 
+ c_3  \bm \alpha \tran (\reallywidehat{\bm \mu}_x \reallywidehat{\bm \mu}_x \tran + \reallywidehat{\bm \Sigma}_x)\bm \alpha,
\end{aligned}
\end{equation}
where
\begin{eqnarray}\nonumber
      c_{1} &= & \frac{g(t_m^*) - g(t_1^*)}{t_m^* - t_1^*} (\reallywidehat{\bm \beta}_1 - \reallywidehat{\bm \beta}_2) (\reallywidehat{\bm \beta}_1 \tran - \reallywidehat{\bm \beta}_2 \tran) \frac{[g(t_m^*)]\tran - [g(t_1^*)]\tran}{t_m^* - t_1^*}, \\\nonumber
c_2 &= &   2 \frac{g(t_m^*) - g(t_1^*)}{t_m^* - t_1^*} (\reallywidehat{\bm \beta}_1 - \reallywidehat{\bm \beta}_2) (\reallywidehat{\bm \Gamma}_1 \tran - \reallywidehat{\bm \Gamma}_2 \tran) \frac{[g(t_m^*)]\tran - [g(t_1^*)]\tran}{t_m^* - t_1^*}, \\\label{catsss}
c_3 &= & \frac{g(t_m^*) - g(t_1^*)}{t_m^* - t_1^*} (\reallywidehat{\bm \Gamma}_1 - \reallywidehat{\bm \Gamma}_2) (\reallywidehat{\bm \Gamma}_1 \tran - \reallywidehat{\bm \Gamma}_2 \tran) \frac{[g(t_m^*)]\tran - [g(t_1^*)]\tran}{t_m^* - t_1^*} .\\\nonumber
\end{eqnarray}
Given a value of $\bm \alpha$, estimates $\reallywidehat{\bm \beta}_k$ and $\reallywidehat{\bm \Gamma}_k$ 
can be obtained by fitting model (\ref{lme2}) (and consequently, the constants $c_1, c_2, c_3,$ are inherently functions of $\bm \alpha$).
Thus, the following algorithm can be used to obtain the estimate $\reallywidehat{\bm \alpha} ^{\text{PATS}}$ in (\ref{catss}) using the Nelder-Mead optimization algorithm \citep{nelder1965simplex}.
\begin{algorithm}[H]
\floatname{algorithm}{Algorithm}
\caption{Estimation of $\bm \alpha$ by maximizing the conditional average tangent slope}
\label{al-ll}
\begin{algorithmic}
\STATE 0: Initialize 
$\reallywidehat{\bm \alpha}^{\text{PATS}(0)}$ 
\STATE 1: Given $\reallywidehat{\bm \alpha}^{\text{PATS}(l)}$, fit the linear mixed-effects model (\ref{lme2}) 
to obtain estimates $\reallywidehat{\bm \beta}_k^{(l)}$ and $\reallywidehat{\bm \Gamma}_k^{(l)}$.
\STATE 2: Use values from step 1 to calculate $\hat c_1^{(l)}, \hat c_2^{(l)}, \hat c_3^{(l)}$ with $\reallywidehat{\bm \beta}_k^{(l)}$ and $\reallywidehat{\bm \Gamma}_k^{(l)}$ in (\ref{catsss}).
\STATE 3: Using the Nelder and Mead algorithm, obtain the update $\reallywidehat{\bm \alpha}^{\text{PATS}(l+1)}$ in (\ref{catss})      
 \STATE 4: Repeat steps 1-3 until convergence.
\end{algorithmic}
\end{algorithm} 
The convergence criterion is to iterate until the cosine similarity between successive estimates, 
$\reallywidehat{\bm \alpha}^{\text{PATS}(l+1)}$ and $\reallywidehat{\bm \alpha}^{\text{PATS}(l)}$, reaches or exceeds 0.99.
In our experience, this algorithm typically achieves convergence within 200 iterations.

\subsubsection{Maximum Likelihood Estimation of Linear Biosignature}\label{mlesection}

In the longitudinal model with a linear link function (\ref{lme2}), primary interest may lay in estimation of the single-index coefficients $\bm{\alpha}$, in which case, the maximum likelihood can be used to estimate $\bm{\alpha}$.
Denote the parameter for the $k$th group as
$\bm \Theta_{k} = (\bm \beta_k, \bm  \Gamma_k, \bm D_k, \sigma_k^2,  \bm{\Theta}_{x})$,
where $\bm{\Theta}_{x}$ parametrizes the distribution of the baseline covariates $\bm{x}$ (assumed consistent across treatment groups).
Assuming normally distributed random effects and error terms (independent of one another),
the outcome for participant $i$ in group $k$ is
$\mu(\bm{x}_{ik}) = 
\bm G_{ik} (\bm \beta_k + \bm \Gamma_k (\bm \alpha ^{\mkern-1.5mu\mathsf{T}} \bm x_{ik}))$ plus the random error. 
The log-likelihood across $K$ groups can be expressed as
\begin{equation} \label{2.3}
\begin{aligned}
  l(\bm \Theta | \bm y, \bm x) = &  \sum_{j=1}^K \sum_{i=1}^{n_k} \Big( \log f_{y|x}(\bm \beta_k,
\bm \Gamma_k, \bm D_k, \bm \alpha, \sigma_k^2|\bm y_{ik}) + \log f_x(\bm x_{ik}|\bm{\Theta}_x) \Big)\\
 = & a + \bm L_1 \bm \alpha + \bm \alpha\tran \bm L_2 \bm \alpha + \sum_{j=1}^K \sum_{i=1}^{n_k} \log f_x(\bm x_{ik}|\bm{\Theta}_{x}),
\end{aligned}
\end{equation}
where, setting $\bm{\Psi}_{ik} = \bm{G}_{ik}\tran \bm{D}_k \bm{G}_{ik}+\sigma^2\bm{I}_{ik}$,
\begin{eqnarray}\nonumber
       a &= & \sum_{j=1}^K \sum_{i=1}^{n_k} \Big(  -\frac{q}{2}\log(2\pi)  - \frac{1}{2}\log(|\bm \Psi_{ik} |) -
\bm y_{ik}\tran \bm \Psi_{ik}^{-1} \bm y_{ik} 
 - \bm \beta_k ^{\mkern-1.5mu\mathsf{T}} \bm G_{ik}\tran \bm \Psi_{ik}^{-1} \bm G_{ik} \bm \beta_k + 2 \bm \beta_k ^{\mkern-1.5mu\mathsf{T}} \bm G_{ik}\tran \bm \Psi_{ik}^{-1} \bm G_{ik} \bm y_{ik} \Big), \\\nonumber
\bm{L}_1 &= &  2 \sum_{j=1}^K \sum_{i=1}^{n_k}\bm \Gamma_k\tran \bm G_{ik}\tran \bm \Psi_{ik}^{-1} (\bm y_{ik} - \bm G_{ik} \bm \beta_k)\bm x_{ik} \tran, \\\label{c1l1l2}
\bm L_2 &= & - \sum_{j=1}^K \sum_{i=1}^{n_k}  \bm \Gamma_k\tran \bm G_{ik}\tran \bm \Psi_{ik}^{-1}\bm G_{ik}  \bm \Gamma_k  \bm x_{ik} \bm x_{ik} \tran,\\\nonumber
\end{eqnarray}
and
$f_x(\bm x_{ik}|\bm{\Theta}_x)$ is the density for the baseline predictors whose parameters are assumed not associated with the biosignature. 
Here, $a\in\mathbb{R}^1$,
and 
$\bm L_1 \sim 1 \times p$ and $\bm L_2 \sim p \times p$ (symmetric and negative definite) are all functions of $\bm{\alpha}$.
The maximum likelihood estimator (MLE) of ${\bm \alpha}$ is then given by 
\begin{equation} \label{ll}
    \reallywidehat{\bm \alpha}^{\text{MLE}} = \argmax_{\bm \alpha}   \text{ } \Big \{a + \bm L_1 \bm \alpha + \bm  \alpha\tran \bm L_2 \bm \alpha \Big \}.
\end{equation}
For a given value $\bm{\alpha} = \bm{\alpha}_0$,  (\ref{ll}) is maximized with respect to $a_1, \bm L_1$, and $\bm L_2$ by
\begin{equation} \label{estll}
    \bm{\alpha}_0^\ast  = \frac{1}{2} \bm L_2 ^{-1} \bm L_1 \tran,
\end{equation}
standardized to have norm equal to 1.  Thus, the following iterative algorithm can 
be used to find $\reallywidehat{\bm \alpha}^{\text{MLE}}$, the MLE  of $\bm{\alpha}$:
\begin{algorithm}[H]
\floatname{algorithm}{Algorithm}
\caption{Estimation of $\bm \alpha$ by maximizing the likelihood function}
\label{al-ll}
\begin{algorithmic}
\STATE 0: Initialize 
$\reallywidehat{\bm \alpha}^{\text{MLE}(0)}$ 
\STATE 1: Given $\reallywidehat{\bm \alpha}^{\text{MLE}(l)}$, fit the linear mixed-effects model (\ref{lme2}) 
to obtain estimates $\reallywidehat{\bm \beta}_k^{(l)}$, $\reallywidehat{\bm \Gamma}_k^{(l)}$ and $\reallywidehat{\bm \Psi}_{ik}^{(l)}$.
\STATE 2: Use values from step 1 to calculate $\reallywidehat{a}_1^{(l)}, \reallywidehat{\bm L}_1^{(l)}, \reallywidehat{\bm L}_2^{(l)}$ with $\reallywidehat{\bm \beta}_k^{(l)}$, $\reallywidehat{\bm \Gamma}_k^{(l)}$ and $\reallywidehat{\bm \Psi}_{ik}^{(l)}$ in (\ref{c1l1l2}).
\STATE 3: Using (\ref{estll}), update the estimate $\reallywidehat{\bm \alpha}^{\text{MLE}(l+1)} \leftarrow
 \frac{1}{2} \hat{\bm{L}}_2^{-1}  (\hat{\bm L}_1^{(l)}) \tran$      
 \STATE 4: Repeat steps 1-3 until convergence.
\end{algorithmic}
\end{algorithm} 
The convergence criterion is to iterate until the cosine similarity between $\reallywidehat{\bm \alpha}^{\text{MLE}(l+1)}$ and $\reallywidehat{\bm \alpha}^{\text{MLE}(l)}$ is equal or larger 
than $0.99$. In our experience, this algorithm usually converges within 20 iterations. 
Note that $\reallywidehat{\bm \alpha}^{\text{MLE}}$ maximizes the likelihood which may not necessarily lead to an optimal ITR.

\subsection{Individualized treatment rules (ITRs)}\label{npats-itr}

Without loss of generality, assume a larger ATS is preferred. 
Then, an ITR based on $\reallywidehat{\bm \alpha}^{\text{NPATS}}$ in Section \ref{est} is defined as 
\begin{equation}\label{non-tdr}
 \reallywidehat{\mathcal{D}}_{\text{NPATS}} (\bm x) =  I \Big( (\reallywidehat{\bm \eta}_2 - \reallywidehat{\bm \eta}_1)\tran \big[\frac{\bm G(t_m^*) - \bm G(t_1^*)}{t_m^* - t_1^*}\otimes \bm a\Big( (\reallywidehat{\bm \alpha}^\text{NPATS}) \tran  \bm x_i\Big)\big] > 0 \Big) + 1.
\end{equation}
Thus, (\ref{non-tdr}) provides an ITR using a trajectory-based single-index model with a flexible link of biosignatures. 
In this paper, cubic B-splines are used for both the time and biosignature links, with $\reallywidehat{\bm{\alpha}}$ estimated from Algorithm \ref{protocol1}. 
Alternatively, the ITRs can be defined using the simpler PATS approach (\ref{ats_k}) using model (\ref{lme2}) or based on the maximum likelihood estimator of $\bm{\alpha}$ from Section \ref{mlesection}, which is specified as:
\begin{equation}\label{tdr_mle_cats}
\reallywidehat{\mathcal{D}}^{\text{PATS/MLE}}
(\bm x) = I \Big ( \frac{\bm g(t_m^*)\tran - \bm g(t_1^*)\tran}{t_m^* - t_1^*} \big(\reallywidehat{\bm \beta}_2 + \reallywidehat{\bm \Gamma}_2 (\reallywidehat{\bm \alpha} \tran \bm x )\big) > \frac{\bm g(t_m^*)\tran - \bm g(t_1^*)\tran}{t_m^* - t_1^*} \big(\reallywidehat{\bm \beta}_1 + \reallywidehat{\bm \Gamma}_1 (\reallywidehat{\bm \alpha} \tran \bm x) \big) \Big) + 1,
\end{equation}

\section{Simulation studies}\label{simsection}

Multiple simulations with $K=2$ treatment groups and different missing data scenarios were conducted to study the performance of the trajectory-based ITRs described above that use longitudinal information based on ${\bm{\alpha}}^{\text{NPATS}}$, ${\bm{\alpha}}^{\text{PATS}}$, and ${\bm{\alpha}}^{\text{MLE}}$. 
Another trajectory-based approach used for comparison is based on a single index embedded into a longitudinal mixed-effects model estimated by maximizing the average Kullback-Leibler Divergence between groups over the predictor space, denoted by {\em LS-KLD} \citep{yao2022single}.

The proposed trajectory-based approaches are compared to other common ITR approaches that use only a simple change score and do not utilize longitudinal information:
\begin{enumerate}
\item[(i)] Single-index model with multiple links (SIMML), a nonparametric single-index approach estimated by maximizing the profile likelihood \citep{park2020single};
\item[(ii)] Linear GEM model estimated under the criterion of maximizing the difference in the treatment-specific slopes, denoted as LinGEM \citep{petkova2017generated};
\item[(iii)] Outcome weighted learning (OWL) \citep{zhao2012estimating}. 
The OWL method is based on a Gaussian radial basis kernel (OWL-Gaussian) implemented in the R package \texttt{DTRlearn}. 
The inverse bandwidth parameter $\sigma_n^2$ in \cite{zhao2012estimating} is chosen from the grid $(0.01, 0.02, 0.04, \ldots, 0.64, 1.28)$, and the tuning parameter $\kappa$ is chosen from the grid $(0.25, 0.5, 1, 2, 4)$ (the default setting of \texttt{DTRlearn}), based on a 5-fold cross-validation;
\item[(iv)] Random forest with 500 trees using the R package \texttt{randomForest} \citep{doubleday2018algorithm};
\item[(v)] Support vector machine (SVM) with the radial kernel, implemented with the R package \texttt{e1071}. 
The parameters in SVM (e.g., $\gamma$ and $C$) are selected with a 5-fold cross-validation \citep{e1071}.
\end{enumerate}

ITR performance for the simulation studies is assessed using the 
value ($V$) of a decision rule defined as the expected mean response when everyone in the population receives treatment according to the rule $\mathcal{D}$, that is 
\begin{equation} \label{value}
    V(\mathcal{D}) = E\Big[E\big[U | \bm x, k =  \mathcal{D}(\bm x) \big]\Big], 
\end{equation}
where $U$ is the extracted scalar outcome.
We evaluated the value $V(\mathcal{D})$ empirically using 
\begin{equation} \label{hatvalue}
    \hat V(\reallywidehat{\mathcal{D}}) = \sum_{i=1}^n U_i \text{I}{ \{A_i = \reallywidehat{\mathcal{D}} (\bm x_i) \} } / \sum_{i=1}^n \text{I}{\{A_i = \reallywidehat{\mathcal{D}}(\bm x_i)\}}
\end{equation}
where $A_i$ is the group assignment and $U_i$ denotes the outcome used for the ITR  for the $i$th subject.  
The performances of ITRs were also compared using the Proportion of Correct Decision (PCD) defined as
\begin{equation} \label{pcd0}
    \text{PCD: }\text{ }\frac{1}{n} \sum_{i=1}^n \text{I} \{ \reallywidehat{\mathcal{D}}(\bm x_i) = \mathcal{D}_0(\bm x_i)\},
\end{equation}
where $ \mathcal{D}_0$ is the true optimal decision which is known in simulation settings. 
Note that the true optimal decision for an individual from simulated data is known since the actual outcomes are generated for both treatments. 
The true optimal treatment is based on the larger of the two simulated scalar outcomes under the two treatments from the fixed and random effect components (e.g., $\bm{b}{ik}$ in (\ref{lme1})), but not the random error (e.g., $\bm{\epsilon}{ik}$ in (\ref{lme1})), since the random error does not correspond to a systematic difference in outcomes due to treatment differences.

\textbf{{Missing data settings: }}
Different scenarios of missingness are considered in the simulations: 
(i) no missing data, i.e., all measures from all subjects are collected; 
(ii) missing completely at random (MCAR), meaning each subject's outcome at each visit (except for the baseline $t=0$) has a 40\% probability of being missing (subjects are independent of each other); 
(iii) missing outcomes due to participants' dropout, which is a monotone missingness independent of outcome.
For the simulations below, the assessment time points are set to $t = 0,1,2,\ldots, 7$, and the proportion of subjects who had no missing data, missed only the last observation, missed the last two, three, and four outcomes were set at 50\%, 30\%, 10\%, 5\%, and 5\%, respectively.

For each scenario, a training data set with a sample size of $n = 200$ ($100$ per group) was generated to estimate the model parameters and the ITRs. 
Independent testing data sets with $n = 1000$ were generated to obtain an independent assessment of the performance of each ITR. 
In all the results presented below, an ITR using the true value of $\bm{\alpha}$ is shown for benchmark comparison. 
For comparison, two na\"ive rule results are also presented: assign all subjects to the control (e.g., placebo) (Allpbo) and assign all subjects to the active treatment (e.g., drug) (Alldrug).

The simulations were conducted in R version 4.3.2 \citep{rrr}. 
The mixed-effect models were fitted with the package \texttt{lme4}, and the Nelder-Mead optimization was implemented using the \texttt{optim} function in R.

\subsection{Simulation Results for Quadratic Trajectories}\label{quadtraj}

\begin{figure}[!p]
\centering\includegraphics[width=\textwidth]{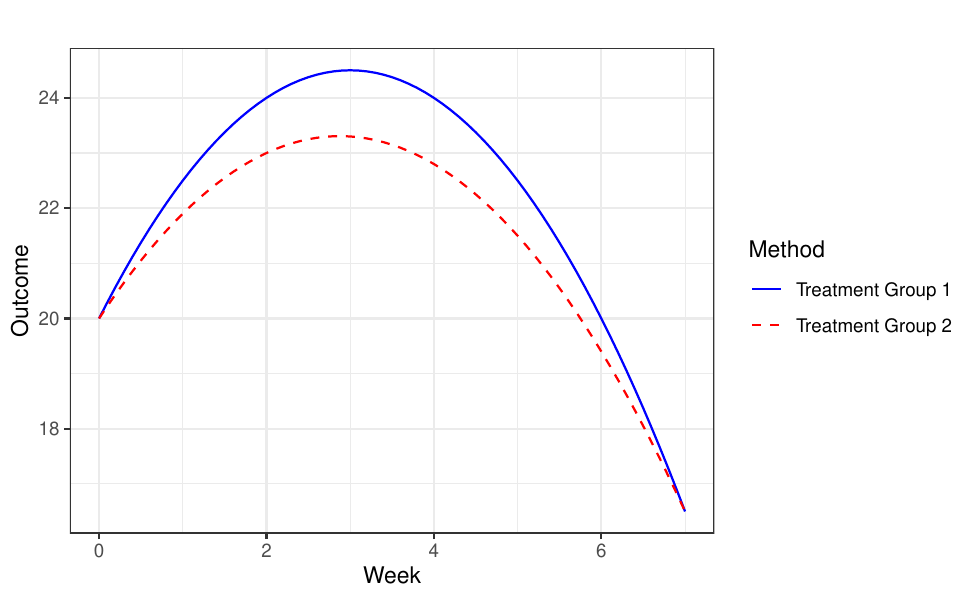}
\caption{\textbf{Average Quadratic Trajectory Plots for the Simulation}.
Quadratic trajectories from treatment groups 1 and 2 were generated based on their average trajectories, which are the blue solid and red dashed curves, respectively.
}
\label{fig:traj}
\end{figure}

For the first simulation illustration, quadratic trajectories were simulated under two treatments. 
Much of the motivation for this work comes from depression and pain studies, which are often of short duration where the outcome trajectories are well-approximated by parabolas. 
Details of the simulation settings for generating the data are in the supplemental information. 
The assessment time points are set to $t = 0,1,2,\ldots, 7$, i.e., the outcomes are collected at baseline and subjects are followed up for 7 weeks.
Figure \ref{fig:traj} shows the mean trajectories used for the quadratic simulation illustration.
It is important to note that since these two mean trajectories are equal at baseline and end-of-study, the ATS for both groups are exactly equal, i.e., on average, there is no difference between mean treatment group trajectories. 
Hence, any treatment difference in terms of ATS for individuals must be due to the interaction component defined by the baseline biosignature $\bm{\alpha} \tran \bm{x}$ represented by $\bm{\Gamma}_k$ from model (\ref{lme2}).
For this simulation illustration, we set
$\bm{\Gamma}_1 = (0, \cos(\theta), \sin(\theta))\tran, \bm \Gamma_2 = (0, \cos(\theta), - \sin(\theta))\tran,$
so that the magnitude of the interaction effect induced by $\bm{\alpha} \tran \bm x$ is determined by $\theta$, the angle between $\bm{\Gamma}_1$ and $\bm{\Gamma}_2$.
The $\theta$'s are chosen to be $0^{\circ}, 1^{\circ},  2^{\circ}, 5^{\circ}$ degrees.  When $\theta=0$, 
$\bm{\Gamma}_1=\bm{\Gamma}_2$
and there is no modifying effect of the baseline covariates on the outcome. 
Larger values of $\theta$ correspond to a greater modifying effect of the linear biosignature on the outcomes.

\begin{figure}[!p]
\centering\includegraphics[width=\textwidth]{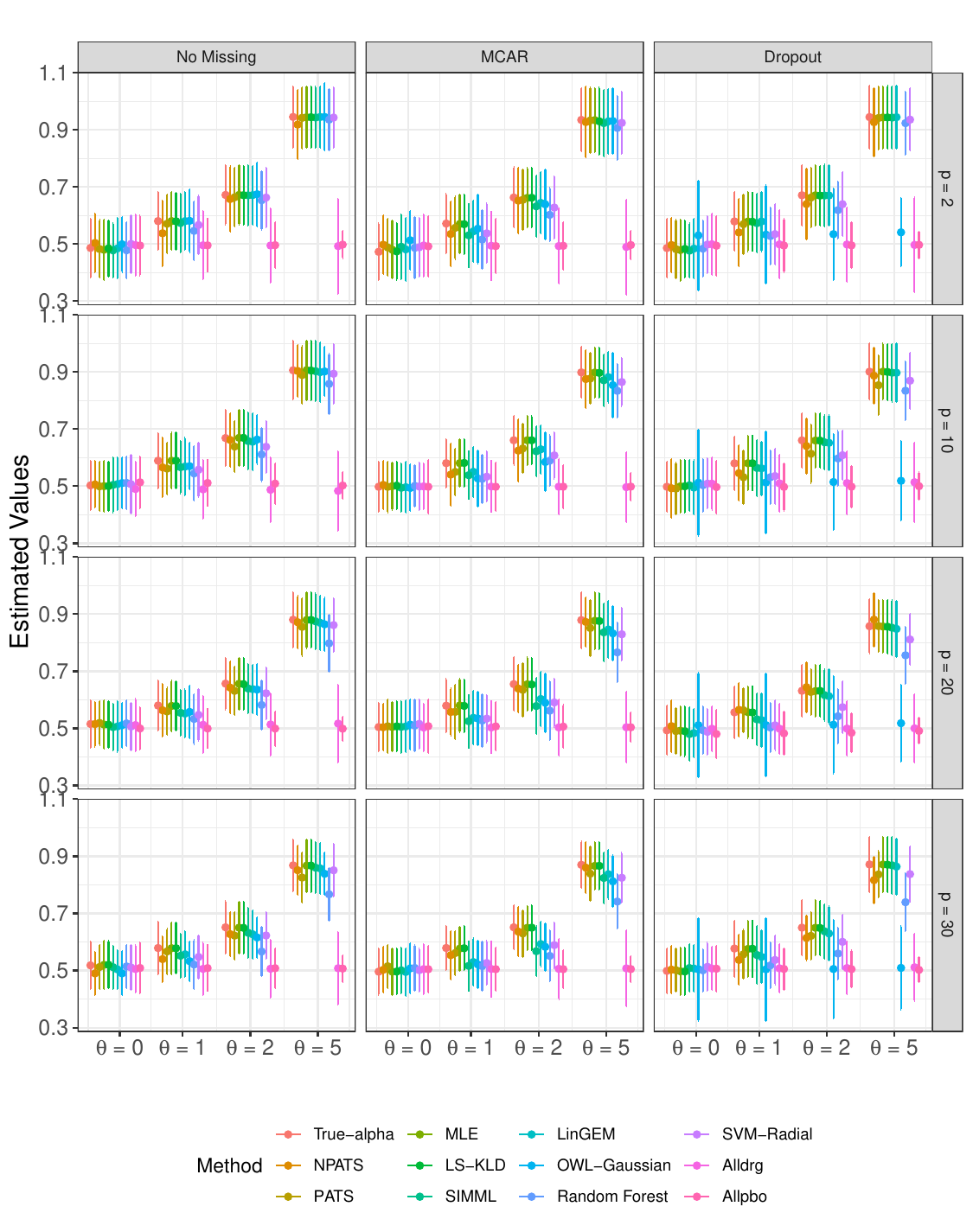}
\caption{
\textbf{Comparisons of the Estimated Values (Quadratic Trajectories}.
Plots of the values estimated from the ITRs obtained from the 200 training data sets.
Each panel corresponds to one of the combinations of $p \in \{2, 10, 20, 30\}$ (across rows) and no missing data, MCAR, or Dropout (across columns). 
The dots represent the mean values estimated across the 200 repetitions and the bars represent the 1.96 standard deviations. 
}
\label{fig:value}
\end{figure}

Figure \ref{fig:value} provides a summary of all simulation results for the quadratic trajectories in terms of the mean value ($\pm$ 1.96 standard deviations) of the ITRs. 
The columns (left to right) correspond to scenarios of no missing data, MCAR, and dropout, respectively; the rows correspond to the number $p$ of predictors ($p = 2, 10, 20, 30$). 
For each panel, the $x$-axis represents the modifying interaction effect in terms of the angle $\theta$ between $\bm{\Gamma}_1$ and $\bm{\Gamma}_2$.

\begin{figure}[htp]
\centering
\includegraphics[width=1\textwidth]{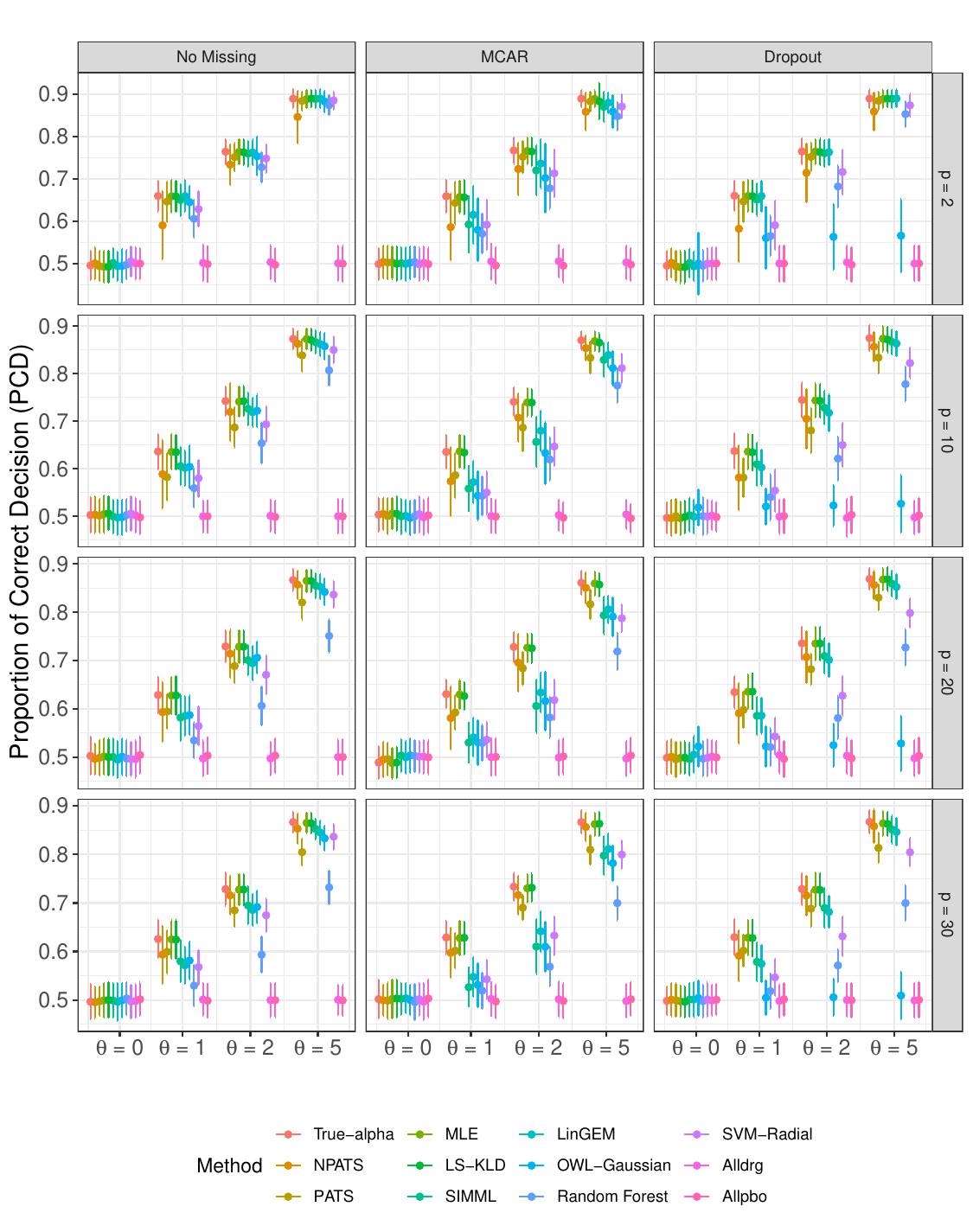}
\caption{
\textbf{Comparisons of the PCD (Quadratic Trajectories).} Plots of the PCD in the quadratic trajectory setting estimated from the treatment decision rules obtained from the 200 training data sets for each of the methods.
Each panel corresponds to one of the combinations of $p \in \{2, 10, 20, 30\}$ and no missing data, MCAR, or Dropout. The dots represent the mean values estimated across the 200 repetitions and the bars represent the 1.96 standard deviations.
}
\label{fig:pcd}
\end{figure}

Uniform treatment assignment methods (Alldrg and Allpbo) serve as benchmarks and are generally surpassed by ITR methods that adapt and perform better as the interaction effects strengthen. 
Without missing data and at a lower predictor count ($p=2$), the methods perform similarly in terms of estimated values. 
As the number of predictors increases ($p=10, 20, 30$), the performances of the different methods diverge, underscoring the impact of model complexity.
Trajectory-based methods like PATS, MLE, and LS-KLD demonstrate remarkable performance consistency across various scenarios, suggesting resilience to missing data. 
All methods perform similarly in the no-interaction setting ($\theta = 0^{\circ}$). 
As $\theta$ increases, performance differences among the methods emerge, with the single-index model approaches generally outperforming SVM and Random Forest for larger $\theta$.
Similar results were observed in terms of PCD, as shown in Figure \ref{fig:pcd}. 
The NPATS method, adding unnecessary complexity, shows reduced performance in quadratic scenarios when the true outcome trajectories are quadratic.

\subsection{Simulation Results for Non-Quadratic Trajectories}\label{simresults-nonquad}

\begin{figure}[!p]
\centering\includegraphics[width=\textwidth]{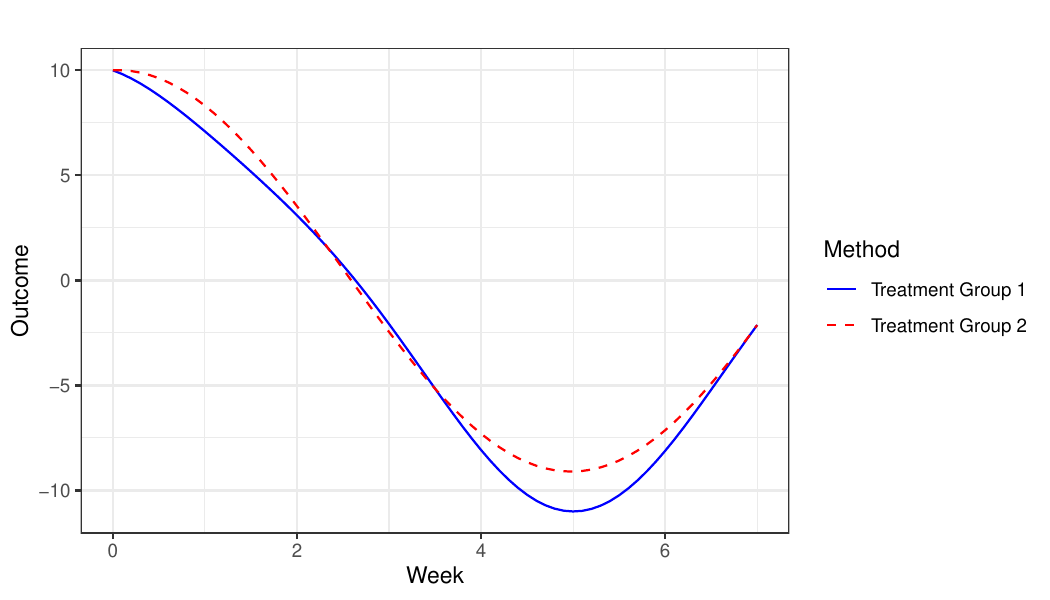}
\caption{\textbf{Nonlinear (Non-Quadratic) Mean Trajectories}. 
The average trajectories for the nonlinear trajectories for the two groups are shown by the blue solid curve and the red dashed curve, respectively.
}
\label{chap3-traj}
\end{figure}

The simulation experiment was repeated for a more complex non-quadratic link function of biosignatures under the same time point and covariate dimensionality settings as in the quadratic trajectory section, with $K = 2$ treatment groups.
The mean trajectory settings are
\begin{equation}
\begin{aligned}
& \text{Treatment Group 1: } \mu_1(t, u) = 
10\cos(\frac{\pi}{5}t) - \sin(\frac{\pi}{2}t) + \sin(\frac{\pi t u}{10});
\\
& \text{Treatment Group 2: } \mu_2(t,u) = 10 \cos(\frac{\pi}{5}t) + \sin(\frac{\pi t}{14}) - \sin(\frac{\pi t u}{10}),
\end{aligned}
\end{equation}
where the interaction term is given by $\sin(\frac{\pi t u}{10})$.
The mean trajectories for both groups when the biosignature $u = 0$ are shown in Figure \ref{chap3-traj}.

Both mean curves show initial improvement followed by deterioration in the later weeks.
The ATS for both mean curves are identical (as was the case for the quadratic illustration), so treatment differences are manifested solely through the moderating effects of baseline covariates. 
The random effects are simulated from a quadratic function of time described in the supplement section. 

The fixed-effect basis functions for the time effect $\bm{g}(t)$ used a cubic B-spline with one knot at $t = 3.5$, and a cubic B-spline with one knot at $u = 0$ was used for the biosignature effect. 
The PATS approach is also illustrated using results from a fitted model with linear and quadratic time trends for both fixed and random effects.
The parameter tuning process for each method is the same as the one in Subsection \ref{quadtraj}.

\begin{figure}[!p]
\centering\includegraphics[width=\textwidth]{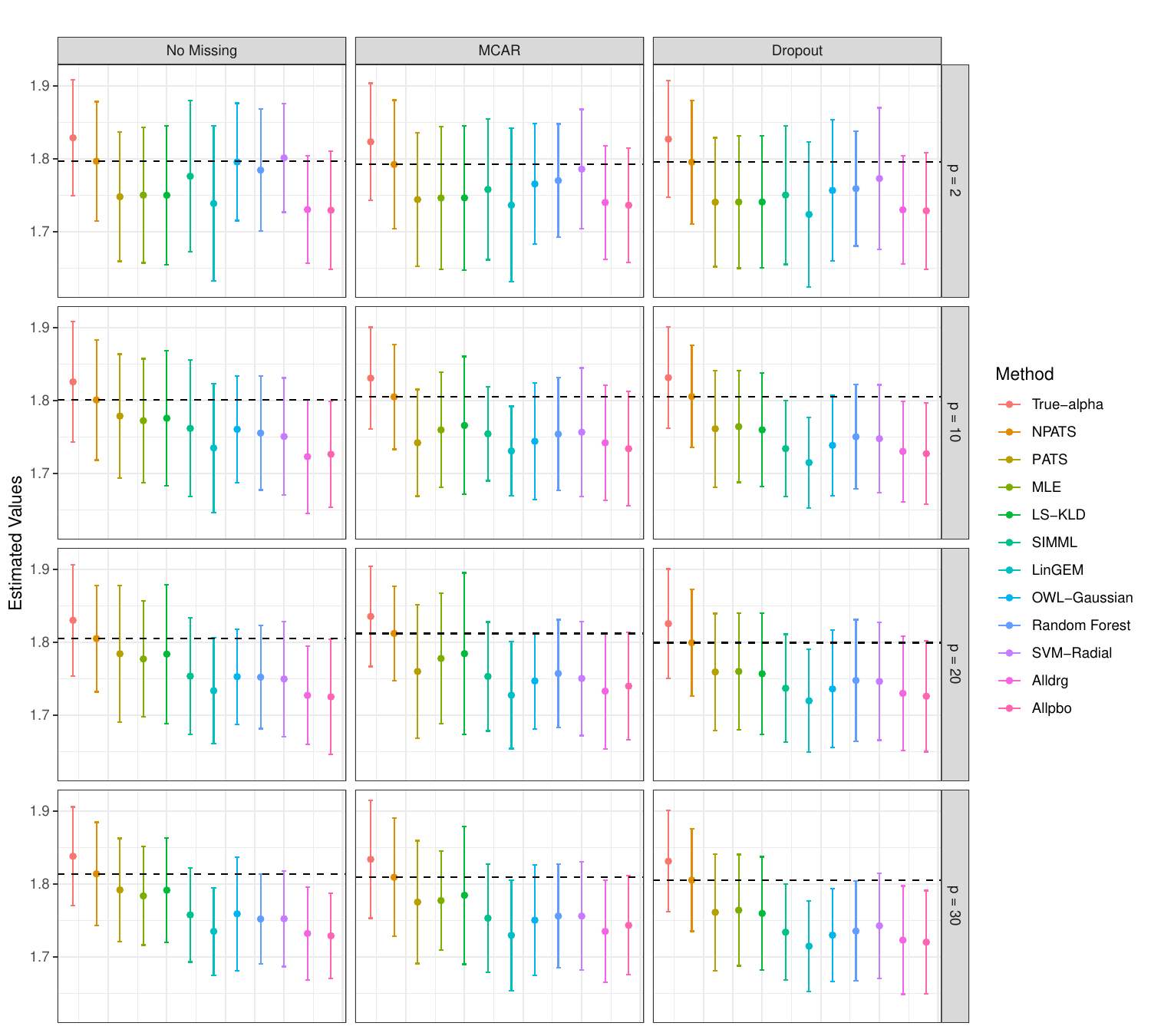}
\caption{\textbf{Comparison of Estimated Values (Nonquadratic Trajectories)}. 
Plots of the values estimated from the ITRs obtained from the 200 training data sets for each of the methods.
Each panel corresponds to one of the combinations of $p \in \{2, 10, 20, 30\}$ and no missing data, MCAR, or Dropout. The dots represent the mean values estimated across the 200 repetitions and the bars represent the 1.96 standard deviations. 
}
\label{chap3-value}
\end{figure}

\begin{figure}[!p]
\centering\includegraphics[width=\textwidth]{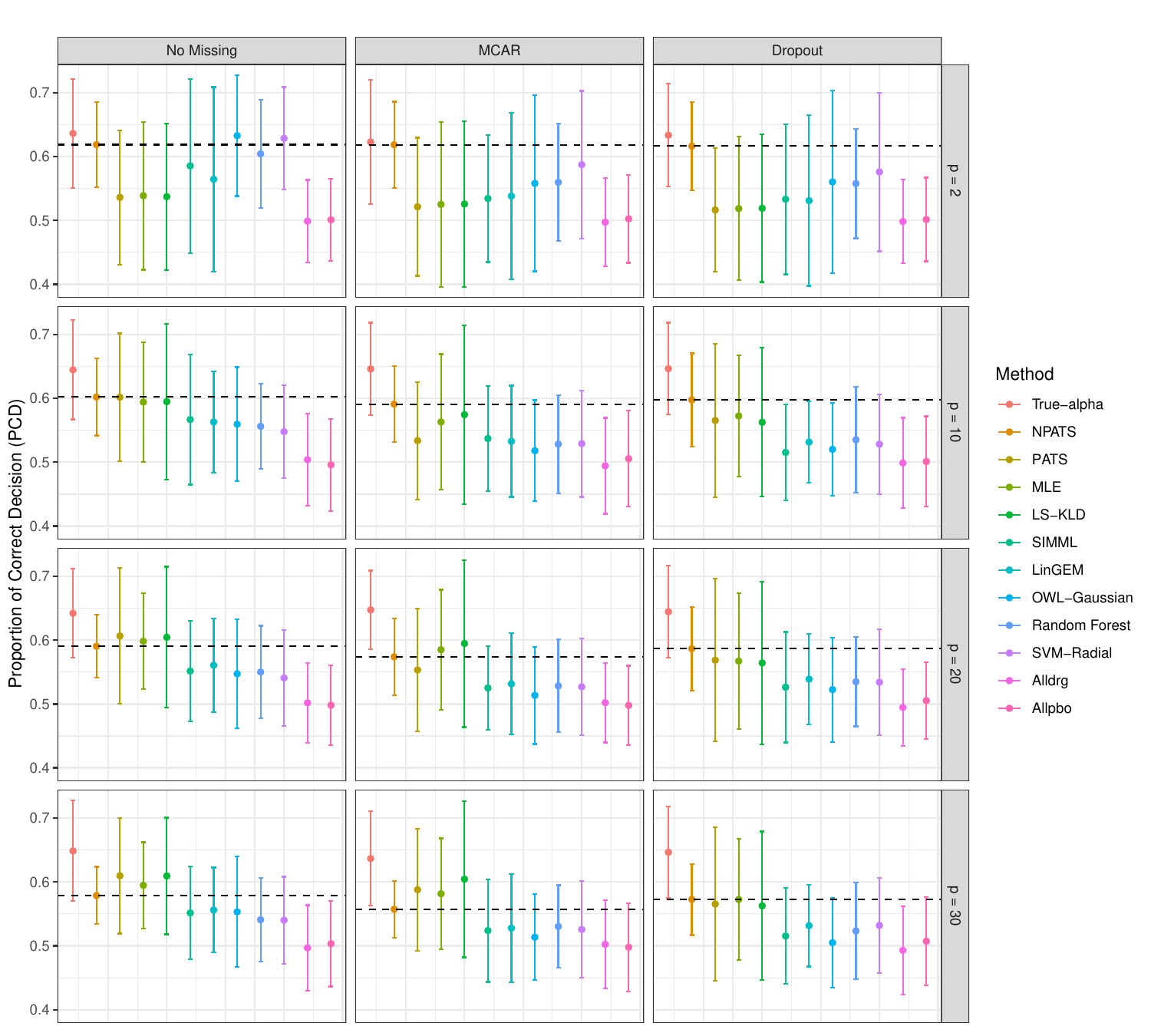}
\caption{\textbf{Comparison of PCD (Nonquadratic Trajectories)}. 
Plots of the proportion of correction decisions (PCD) from the 200 training data sets for each of the methods for the nonlinear (non-quadratic) trajectories.
Each panel corresponds to one of the combinations of $p \in \{2, 10, 20, 30\}$ and no missing data, MCAR, or Dropout. The dots represent the mean values estimated across the 200 repetitions and the bars represent the 1.96 standard deviations.
}
\label{chap3-pcd}
\end{figure}

Figures \ref{chap3-value} and \ref{chap3-pcd}.
show results in terms of value and PCD respectively.
The horizontal dashed line marks the average value assessed for the proposed approach, NPATS. 
For comparison, results are shown using the true $\bm{\alpha}$ that generated the data (``True-alpha'').
NPATS tended to perform best compared to all other approaches across all scenarios, particularly in the missing data scenarios. 
The three trajectory-based single-index approaches, PATS, MLE, and LS-KLD, tended to perform similarly in all scenarios and better than methods that do not utilize longitudinal information.

\section{Application to a depression study}\label{embarc}

The ITRs developed above are applied in this section to data from the EMBARC longitudinal depression trial \citep{trivedi2016establishing} to illustrate the utility of the trajectory-based ITRs. 
The primary outcome, the Hamilton Depression Rating Scale (HDRS), where lower scores correspond to less depression, was evaluated for each participant at weeks 0 (baseline), 1, 2, 3, 4, 6, and 8. 
Of the 160 patients, 87 were randomized to placebo and 73 to an active antidepressant drug treatment.

To compare various ITRs for the EMBARC data, empirical estimates of the value of the ITRs were obtained using the {\em inverse probability weighted estimator} (IPWE) ~\citep{murphy2005generalization}:
\begin{equation}\label{ipwe}
    \text{IPWE} = \sum_{i = 1}^n \{I(A_i = \hat A_i) U_i \}/ \sum_{i = 1}^n I(A_i = \hat A_i),
\end{equation}
where
$A_i \in \{1,2\}$ is the observed treatment assignment and $\hat{A}_i$ is the treatment assignment based on the estimated ITR  (1 = active treatment, 2 = placebo);
$U_i$ is the change score, i.e., the difference between the first and last HDRS outcomes. 
Change score is being used here to compare ITRs since it is a measure that is extractable from all methods under consideration.
Larger values of $U_i$ correspond to more improvement from baseline.

As an initial illustration using this data, univariate analyses were conducted for 
the following $p=10$ baseline predictors to motivate the use of single-index ITRs:
$(x_1)$ Age at Evaluation,
 $(x_2)$ Anger attack,
 $(x_3)$ Axis II,
$(x_4)$ Chronicity,
$(x_5)$ Hypersomnia, $(x_6)$ Sex, 
$(x_7)$ Number of correct responses for ``A not B'' Working memory task,
$(x_8)$ Median Reaction time,
$(x_9)$ Flanker Reaction Time,
$(x_{10})$ Flanker Accuracy.
If separate linear regression models are fit for each of these baseline covariates, with HDRS as the outcome modeled by the predictor, the treatment indicator, and their interaction, then the only predictors producing significant interaction effects (e.g., $p<0.05$) are ``Age at Evaluation" ($p=0.006$) and ``Flanker Accuracy" ($p=0.005$).
The treatment-modifying effect for each baseline characteristic was also evaluated using the IPWE (\ref{ipwe}).
Simple linear regressions and B-spline regressions of the HDRS improvement on each covariate were obtained separately in the antidepressant and placebo groups. 
A 10-fold cross-validation (CV) was conducted to assess the ITR performances, and this 10-fold CV was repeated for 100 random splits of the data, then averaged over all random splits.
The CV results from applying linear regressions (Linear) and B-spline regressions (Nonpar.) are summarized in Table \ref{tab2:2}, where linear models or B-spline models were applied separately to each covariate to model their relationship with the outcome change score.

\begin{table}[!p]
\centering
\caption{Demographic Characteristics and IPWE Values for Baseline Covariates\label{tab2:2}. The second and third columns summarize the mean (SD) for continuous variables and the counts ($\%$) for categorical variables in each group.
``Number of correct responses" represents the number of correct responses in the ``A not B” Working memory task.
The $p$-values for the interaction effects between outcomes and each covariate are provided in the fourth column.
LRTs were used to calculate the $p-$values. 
The IPWE values obtained by fitting linear regressions or B-spline regressions are shown in the last two columns.}
\label{Table1}
{\tabcolsep=1.25pt
\begin{tabular}{@{\extracolsep\fill}lcccccc@{\extracolsep\fill}}
\toprule
\textbf{Baseline Covariates} & \textbf{Treatment} & \textbf{Placebo} & \textbf{$p$-value} & \textbf{Linear} & \textbf{Nonpar.} \\
\midrule
$(x_1)$ Age at Evaluation & 37.44 (15.01) & 38.30 (13.03) & 0.006 & 5.71 & 5.13 \\
$(x_2)$ Anger Attack & 3.08 (2.16) & 2.97 (2.07) & 0.839 & 6.66 & 7.23 \\
$(x_3)$ Axis II & 3.92 (1.40) & 3.91 (1.50) & 0.410 & 6.56 & 7.20 \\
$(x_4)$ Chronicity & & & 0.219 & 7.08 & 7.08 \\
$~$ $~$ $~$ $~$ Yes & 39 (53.4) & 47 (54.0) & & & \\
$~$ $~$ $~$ $~$ No & 34 (46.6) & 40 (46.0) & & & \\
$(x_5)$ Hypersomnia & & & 0.667 & 6.33 & 6.33 \\
$~$ $~$ $~$ $~$ Yes & 17 (23.3) & 16 (18.4) & & & \\
$~$ $~$ $~$ $~$ No & 56 (76.7) & 71 (81.6) & & & \\
$(x_6)$ Sex & & & 0.921 & 6.36 & 6.36 \\
$~$ $~$ $~$ $~$ Female & 26 (35.6) & 32 (36.8) & & & \\
$~$ $~$ $~$ $~$ Male & 47 (64.4) & 55 (63.2) & & & \\
$(x_7)$ Number of Correct Responses & 0.05 (0.82) & 0.23 (0.71) & 0.309 & 6.46 & 6.55 \\
$(x_8)$ Median Reaction Time & 0.51 (1.78) & 0.01 (1.10) & 0.649 & 6.60 & 6.27 \\
$(x_9)$ Flanker Reaction Time & 61.36 (28.61) & 59.52 (24.40) & 0.075 & 5.65 & 5.77 \\
$(x_{10})$ Flanker Accuracy & 0.22 (0.15) & 0.22 (0.15) & 0.005 & 5.44 & 6.00 \\
\bottomrule
\end{tabular}}
\end{table}

\begin{figure}[!p]
\centering\includegraphics[width=\textwidth]{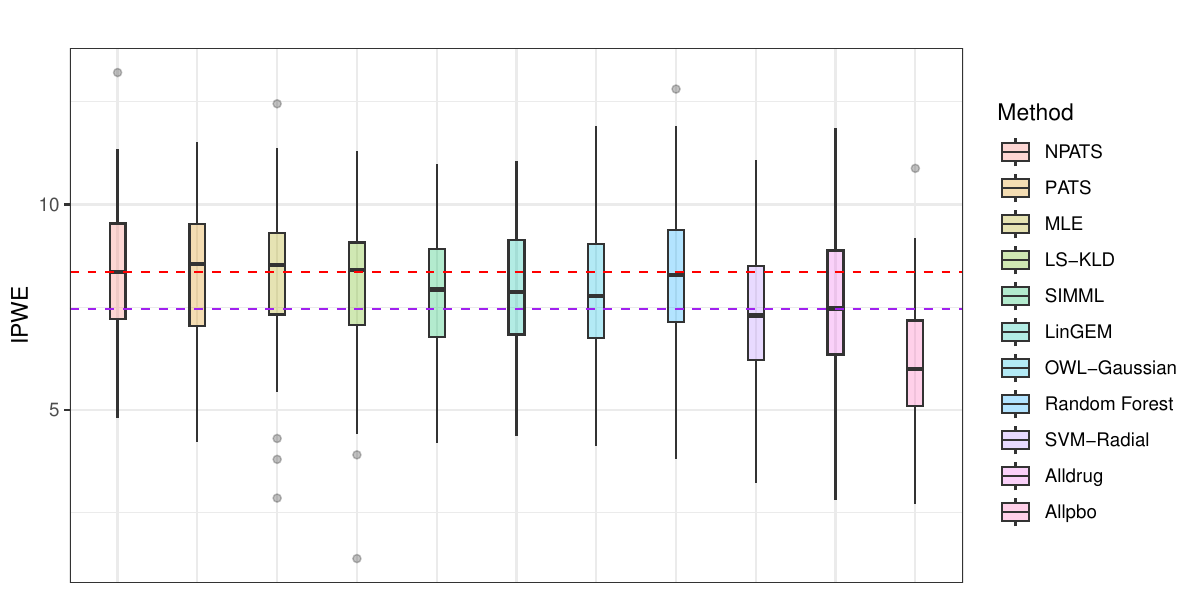}
\caption{\textbf{Evaluation of the IPWE}. 
Box plots of the estimated IPWE values across the 1000 splits (higher values are preferred). The median and standard deviations for each method are:
NPATS: 8.35 (1.61);
PATS: 8.55 (1.68);
MLE:  8.53 (1.64);
LS-KLD: 8.41 (1.69); 
SIMML: 7.94 (1.48);
LinGEM: 7.87 (1.58);
OWL-Gaussian: 7.84 (1.67);
Random Forest: 8.29 (1.64); 
SVM-Radial: 7.30 (1.55);
Alldrug: 7.47 (1.7); 
Allpbo: 6.00 (1.52).
}
\label{ipwebox}
\end{figure}

Since no single predictor appears to be a strong effect modifier by itself, the ITRs using multiple predictors were investigated and compared using IPWE 
and a 10-fold CV with 100 random splits as was done for models with a single predictor.
The IPWEs for the ITRs are summarized in Figure \ref{ipwebox}.
The purple horizontal dashed line shows the median IPWE value across the 1000 repetitions (10-fold times 100 random splits) when all patients are assigned to the active treatment group, while the red dashed line represents the median IPWE value of the NPATS.  

Figure \ref{ipwebox} shows that the trajectory-based methods generally offer comparable performances among themselves, and they all perform better than methods that do not
incorporate longitudinal information. 
Random Forest performs competitively exhibiting a median IPWE value that is on par with the trajectory-based methods whereas the SVM approach performs quite poorly.
Also note that all the methods shown in Figure \ref{ipwebox} outperformed the single predictor models shown in Table \ref{tab2:2} in terms of value.

\subsection{ITR Comparisons Across Multiple Covariate Combinations}\label{additional}

\begin{figure}[!p]
\centering\includegraphics[width=\textwidth]{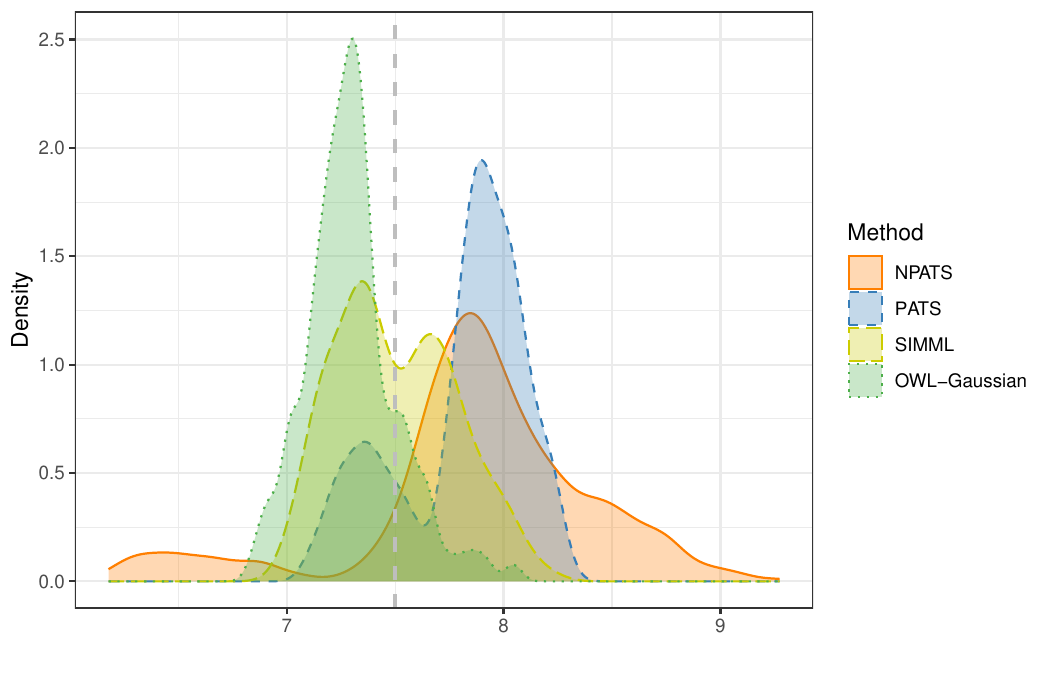}
\caption{\textbf{Density Plots of the IPWE Estimations}. 
The density plots of the mean {CV} IPWE values across 500 combination of baseline covariates. The vertical dashed line {re}presents the IPWE value by assigning all subjects to the drug group. The mean (SD) of each method is: 
NPATS: 7.89 (0.55)
PATS: 7.81 (0.29);
SIMML: 7.38 (0.27);
OWL-Gaussian: 7.30 (0.21).}
\label{d}
\end{figure}

The performances of the ITRs depend on the choice of predictors included in the model. 
To obtain an unbiased assessment comparing the different ITRs, we examined ITR performance for 500 models with various combinations of baseline covariates. 
These 500 models included random combinations of predictors with varying numbers of predictors. 
For this illustration, a total of 21 baseline covariates from demographic, clinical, and behavioral phenotyping data were used.
For the $i$th combination, the number of covariates $n_{ci}$ was chosen at random from a uniform distribution on the set $\{8,9,\dots, 21\}$. Then $n_{ci}$ covariates were randomly selected from the demographic, clinical, and behavior phenotyping covariates sets.
Since ``Age at Evaluation" and ``Flanker Accuracy" have significant interactions with the outcome, each covariate combination included these two covariates, as well as "Flanker Reaction Time", which is an important measure in clinical practice \citep{dillon2015computational}.
One hundred repetitions of 10-fold CV were conducted, and the mean IPWE values were calculated across the 1,000 training sets. 
Figure \ref{d} shows density plots of the IPWEs estimated by various ITR approaches (NPATS, PATS, SIMML, and OWL-Gaussian) across 500 baseline covariate combinations. 
To ensure clarity, we omitted the results for SVM and Random Forest, as their performance and estimated IPWE were very similar to those of Alldrug. 
We also excluded LinGEM results because they were very close to SIMML. 
Additionally, we only included the best-performing parametric trajectory-based method, PATS, omitting the results of MLE and LS-KLD.

By simply allocating all individuals to the active treatment group (Alldrug), the IPWE value is 7.49 (shown as the vertical dashed line). 
OWL-Gaussian performs the worst on average, while SIMML does somewhat better.
Both SIMML and OWL-Gaussian have smaller IPWEs on average than simply assigning all subjects to the active treatment group. 
NPATS has the highest mean IPWE across 500 combinations of baseline covariates, followed by PATS. 
The ITRs based on MLE and LS-KLD performed worse than NPATS; their results are not shown in Figure \ref{d}. 
The performance of NPATS across covariate combinations is more variable than that of PATS, indicating that some covariate combinations led to poor performance of NPATS.
Note that in this analysis, there are combinations of predictors where the ITRs based on a single index perform worse than the Alldrug policy. 
This is expected because some models may include variables with negligible moderator effects, and no combination of these variables may yield a strong ITR. 
The predictors ``Age at Evaluation" and ``Flanker Accuracy" were included in all combinations generated, but similar results were obtained without this restriction.
In conclusion, the single-index ITRs NPATS and PATS, which take into account longitudinal effects, have better performance for individualized treatment decision rules for the EMBARC data, especially compared to ITR approaches that ignore the longitudinal information in the data.

\section{Discussion}\label{discussion}

Most clinical trials collect outcome data repeatedly over time, but ITR approaches in precision medicine usually ignore the information available from the longitudinal nature of the data. 
This paper has developed trajectory-based ITRs that can embed linear combinations of baseline predictors (i.e., biosignatures) into longitudinal models to improve ITR performance.
Smooth nonparametric (NPATS) and parametric quadratic-based (PATS, MLE) ITRs have been developed, studied in simulations, and applied to a depression clinical trial. 
Often, such trials are of short duration, in which case the parametric approach to trajectory-based ITR construction may be expected to perform better than NPATS, since relatively simple trajectory shapes (e.g., parabolas) are seen in many of these trials.

Extensions for future investigations into the proposed trajectory-based ITRs include accommodating trajectories that exhibit a flattening out (e.g., horizontal asymptote), which are often observed in clinical trials. 
Additionally, introducing penalty terms into the trajectory-based models to penalize over-smoothing or an $L^1$-type penalty to implement variable selection while incorporating longitudinal information would be valuable.

\section*{Acknowledgments}

This work was supported by the National Institute of Mental Health (NIMH) grant 5 R01 MH099003. 
We also appreciate the help and support from Drs. Hyung Park, Mengling Liu, Samprit Banerjee, Binhuan Wang, and Jiyuan Hu.

\section{Supplemental Information}\label{supplement}

This section provides specific information on the parameter settings used in the simulation studies in Section \ref{simsection}.

{\bf Covariates:}
The number of covariates $p$ used in the simulation scenarios were $p = 2, 10, 20, 30$. 
We set $\bm \alpha = (1,..,p)\tran$, standardized to have norm one.  
The baseline covariates $\bm x_{ik}$ were sampled from $MVN(\bm \mu_x, \bm \Sigma_x)$, where $\bm \Sigma_x$ is a $p \times p$ matrix and has 1's on the diagonal and $0.5^{|i-j|}$ for the $(i-j)$th element. 
The vector of mean values, $\bm \mu_x$, is set as $\bm \mu_x = \big(-p, -(p-1),...,2,1\big)$, i.e., half are negative and the other half are positive, which makes the biosignature $\bm \alpha \tran \bm x$ have mean $0$.

\subsection{Quadratic Trajectories}\label{quadsupplement}

For the simulated quadratic trajectories, the {assessment time points are set to} $ t = 0,1,2,..., 7$, i.e., the outcomes are collected at baseline and subjects are followed up for 7 weeks. 
The fixed-effect coefficient parameters $\bm \beta_1$ and $\bm \beta_2$ for treatment groups 1 and 2, are chosen to be similar to each other (as was the case {for the estimated} EMBARC parameters), 
$$\bm{\beta}_1 = (20, 3, -0.5)\tran \;\;   \mbox{{and}} \;\;  \bm{\beta}_2 = (20,2.3,-0.4)\tran .$$
The vectors of random effects are set as $$\bm{b}_{i,1} \sim MVN(\bm 0, \bm D_1) \text{ and } \bm b_{i,2} \sim MVN(\bm 0, \bm D_2)$$ with 
$$\bm D_{1} =  \left(\begin{array}
{ccc}
0.5 & -0.1 & -0.01\\
-0.1 & 0.5 & -0.01 \\
-0.01 & -0.01 & 0.01
\end{array}\right) \;\; \mbox{and} \;\; \bm D_2 = \left(\begin{array}
{ccc}
0.5 & -0.12 & -0.01 \\
-0.12 & 0.5 & -0.01 \\
-0.01 & - 0.01 & 0.01
\end{array}\right).$$

For the nonlinear simulation setting in Section (\ref{simresults-nonquad}), the random effects used a quadratic effect by letting
$$\bm Z_{ik} = \left(\begin{array}
{ccc}
1 & t_{i1k} & t_{i1k}^2 \\
\vdots & \vdots & \vdots \\
1 & t_{im_{ik}k} &  t_{im_{ik}k}^2 \\
\end{array}\right),$$
and $\bm b_{ik}$ simulated from MVN with mean zero and covariance matrices
$$\bm D_{1} = \bm D_{2} =  \left(\begin{array}
{ccc}
0.5 & -0.1 & -0.01\\
-0.1 & 0.5 & -0.01 \\
-0.01 & -0.01 & 0.01
\end{array}\right)$$
for both groups.
Random errors $\epsilon_{ijk}$s were generated similar to the quadratic trajectory scenario.

\subsection{Data and Code Availability}
\label{sec5}

The EMBARC data and additional corresponding sample codes can be obtained upon request from the corresponding author at \href{ly1192@nyu.edu}{ly1192@nyu.edu}.
Sample simulation codes are available on \href{https://github.com/sakuramomo1005/Trajectory_based_ITR}{https://github.com/sakuramomo1005/Trajectory$\_$based\_ITR}.


\bibliographystyle{Chicago}

\end{document}